\documentclass[letterpaper,twocolumn,showpacs,pra,aps]{revtex4-1}

\usepackage{amsmath}  
\usepackage{amsfonts} 
\usepackage{graphicx} 
\usepackage{amssymb}

\begin{document}

\title{Screening and plasma oscillations in an electron gas in the hydrodynamic approximation}

\author{Eugene B. Kolomeisky$^{1}$ and Joseph P. Straley$^{2}$}

\affiliation
{$^{1}$Department of Physics, University of Virginia, P. O. Box 400714,
Charlottesville, Virginia 22904-4714, USA\\
$^{2}$Department of Physics and Astronomy, University of Kentucky,
Lexington, Kentucky 40506-0055, USA}

\date{\today}

\begin{abstract}
A hydrodynamic theory of screening in a generic electron gas of arbitrary dimensionality is given that encompasses all previously studied cases and clarifies the predictions of the many-body approach.  We find that long-wavelength plasma oscillations are classical phenomena with quantum-mechanical effects playing no explicit role.  The character of the oscillations is solely dictated by the dimensionality of the electron system and its equation of state in the neutral limit.  Materials whose excitations are described by the Dirac dispersion law -- such as doped graphene or a Weyl semimetal -- are no exception to this rule.          
\end{abstract}


\maketitle

\section{Motivation}

An interacting electron gas in the presence of a uniform positively charged background (the jellium model) is one of the paradigms that has shaped our understanding of the physics of bulk metals, doped semiconductors \cite{Bohm_Pines,Pines_Nozieres,Mahan} and of various two-dimensional systems \cite{AFS}.  It provides a reasonable approximation to real materials, correctly describing the ground-state properties,  screening, and the excitation spectrum.  In the long-wavelength limit, the excitations have the character of classical plasma oscillations \cite{Tonks},  implying that their long-wavelength properties can be derived from a macroscopic theory \cite{Jackson}. 

In the usual electron gases the excitation energy is quadratic in the wave vector ($E(\textbf{q})=\hbar^{2}q^{2}/2m$, where $q=|\textbf{q}|$ and $m$ is the effective electron mass). For graphene, however, the low energy excitations obey the Dirac dispersion law
\begin{equation}
\label{dispersion_law}
E(\textbf{q})=\hbar v_{F}q
\end{equation}
where $v_{F}\approx c/300$ is the limiting (Fermi) velocity \cite{graphene_review}.   There also are three-dimensional "Weyl materials" that exhibit this dispersion relation \cite{AB}.

The idea that long-wavelength plasmons correspond to classical plasma oscillations is challenged by the example of graphene: while the functional dependence of the plasma frequency on the wave vector $\Omega(\textbf{q})$ agrees with classical expectations, the spectrum features an explicit dependence on Planck's constant $\hbar$ and a non-classical dependence of the plasma frequency on doping \cite{Hwang_D_Sarma}.  Non-classical behavior is also predicted whenever the excitations obey Eq.(\ref{dispersion_law}), implying that the plasma oscillations of a Dirac plasma (system of electrons or holes obeying the dispersion law (\ref{dispersion_law}) in the presence of neutralizing background) can only be understood as a quantum effect, even in the long wavelength limit \cite{DasHwang}.    

In this paper we will resolve the conceptual puzzle of non-classical behavior discovered in Refs.\cite{Hwang_D_Sarma,DasHwang}.  Despite the appearance of $\hbar$ in the plasma frequency and the non-classical dependence on doping, the long-wavelength properties of plasma oscillations will be derived from hydrodynamics;  the microscopic nature of the electron plasma only enters through the material parameters of the theory.     

Our analysis is modeled after Bloch's hydrodynamic generalization of the semiclassical Thomas-Fermi (TF) model of a neutral atom \cite{Bloch}.  The application Bloch specifically had in mind was to explain the stopping power.  Later Bloch's equations were employed to describe a small metal particle represented by an electron gas confined within a sphere \cite{Jensen}, and photoabsorption and the collective charge oscillations of a TF atom \cite{Ball}.  A formulation similar to Bloch's  has been given by Fetter \cite{Fetter} who related the hydrodynamic and many-body predictions regarding screening and plasmons in three- and two-dimensional electron gases with parabolic dispersion laws.  Our analysis recovers Fetter's findings as special cases.     

To be definite we focus on a $d$-dimensional gas of electrons (the theory for the holes is the same) in the presence of a neutralizing background charge of uniform number density $n_{d}$;  the practically relevant cases are a bulk system ($d=3$), a layer ($d=2$), and a wire ($d=1$).  We start by summarizing the static long-wavelength screening properties of the electron gas, followed by the hydrodynamic generalization of the theory for  electrons obeying the parabolic dispersion law, and conclude with an even more general theory which accommodates the Dirac dispersion law (\ref{dispersion_law});  this is where our central results lie. 

\section{Static screening}
  
The total potential $\varphi(\textbf{r})$ felt by an electron of charge $e$ at position $\textbf{r}$ is due to the external potential $\varphi_{ext}(\textbf{r})$ and to the potential caused by the net local charge of the remaining electrons of number density $n(\textbf{r})$ and neutralizing background of density $n_{d}$:     
\begin{equation}
\label{scpotential}
\varphi(\textbf{r})=\varphi_{ext}(\textbf{r})+\frac{e}{\kappa}\int \frac{[n(\textbf{r}')-n_{d}]}{|\textbf{r}-\textbf{r}'|}d^{d}r'
\end{equation}
where $\kappa$ is the background dielectric constant.  In thermodynamic equilibrium the electrochemical potential
\begin{equation}
\label{equilibrium}
\mu=\zeta(n)+e\varphi(\textbf{r})
\end{equation}
is fixed at a constant value $\zeta(n_{d})$ so that $n=n_{d}$ and $\varphi=0$ (for $\varphi_{ext}=0$).  Here $\zeta(n)$ is the chemical potential of the electrons (whose temperature dependence is for brevity not displayed) in the absence of the perturbing potential $\varphi(\textbf{r})$.  In the presence of a weak external potential $\varphi_{ext}, $ Eqs.(\ref{scpotential}) and (\ref{equilibrium}) (when $\mu=\zeta(n_{d})$) can be linearized about $n=n_{d}$ and $\varphi=0$. In terms of $n(\textbf{q})$, $\varphi_{ext}(\textbf{q})$, and $\varphi(\textbf{q})$, the Fourier transforms of $\delta n(\textbf{r})=n(\textbf{r})-n_{d}$, $\varphi_{ext}(\textbf{r})$, and $\varphi(\textbf{r})$, respectively, the outcome can be written as
\begin{equation}
\label{potential_Fourier}
\varphi(\textbf{q})=\varphi_{ext}(\textbf{q})+\frac{e}{\kappa}n(\textbf{q})f_{d}(q),
\end{equation}    
\begin{equation}
\label{equilibrium_modified}
\frac{\partial\zeta}{\partial n_{d}}n(\textbf{q})+e\varphi(\textbf{q})=0
\end{equation}
where $f_{d}(q\rightarrow 0)\simeq q^{1-d}$ is the Fourier transform of $1/r$, which for the relevant cases is given by
\begin{equation}
\label{Fourier_of_Coulomb}
f_{1}(q)\approx2\ln\frac{1}{qa},qa\ll 1,~~f_{2}(q)=\frac{2\pi}{q},~~f_{3}(q)=\frac{4\pi}{q^{2}}
\end{equation} 
where $a$ is the wire radius.  Elimination of $n(\textbf{q})$ from Eqs.(\ref{potential_Fourier}) and (\ref{equilibrium_modified}) establishes that $\varphi(\textbf{q})=\varphi_{ext}(\textbf{q})/\varepsilon_{d}(\textbf{q})$, thus giving the static dielectric function:
\begin{equation}
\label{phi_vs_phi_ext}
\varepsilon_{d}(\textbf{q})=1+\frac{e^{2}}{\kappa}\frac{\partial n_{d}}{\partial \zeta}f_{d}(q)
\end{equation}
which is determined by the thermodynamic density of states $\partial n_{d}/\partial \zeta$ and the dimensionality of the electron system which enters through $f_{d}(q)$. The $d\neq 1$ expression for the dielectric function can be equivalently rewritten as
\begin{equation}
\label{static_diel_function}
\varepsilon_{d}(\textbf{q})=1+\left (\frac{q_{s}}{q}\right )^{d-1}, q_{s}\simeq \left (\frac{e^{2}}{\kappa}\frac{\partial n_{d}}{\partial \zeta}\right )^{1/(d-1)}
\end{equation}
where $q_{s}^{-1}$ is the Debye screening radius of the electron gas.  Long-wavelength ($q \ll q_{s}$) perturbations are completely screened, while their short-wavelength counterparts ($q \gg q_{s}$) are unaffected.  For $d=3$ or $d=2$ Eq.(\ref{static_diel_function}) reproduces the well-known results \cite{AFS,Pines_Nozieres}.  In the $d=1$ case one has $\varepsilon_{1}(\textbf{q})\approx1+(2e^{2}/\kappa)(\partial n_{1}/\partial \zeta)\ln(1/qa)$.

\section{Dynamical screening}

When the electrochemical potential is not constant across the system there will be a net force exerted on the electrons \begin{equation}
\label{driving_force}
\textbf{F}=-\nabla \mu=-\nabla (\zeta + e\varphi)
\end{equation}
causing them to move (in layer or wire geometry, there are also confining forces normal to the surfaces constraining motion in those directions; the effect is to reduce the dimensionality of the differential operators here and below).   Following Bloch \cite{Bloch} this is described by treating the electrons as charged ideal liquid characterized by the local position- and time-dependent number density $n(\textbf{r},t)$ and velocity $\textbf{u}(\textbf{r},t)$ fields, which are related by the continuity equation
\begin{equation}
\label{continuity}
\frac{\partial n}{\partial t}+\nabla \cdot(n\textbf{u})=0
\end{equation}
Since the electron velocities are significantly slower than the speed of light, the effects of retardation are neglected from the outset so that Eqs.(\ref{scpotential}) and (\ref{potential_Fourier}) continue to hold except that the potentials and density acquire time dependence.  

\subsection{Parabolic dispersion law}

When the underlying particles of the liquid exhibit a parabolic dispersion law $E(\textbf{q})=\hbar^{2}q^{2}/2m$, the equation of motion is Newton's second law $md\textbf{u}/dt=\textbf{F}$ or
\begin{equation}
\label{2nd_law}
m\left (\frac{\partial \textbf{u}}{\partial t} +(\textbf{u}\cdot \nabla)\textbf{u}\right )=-\nabla (\zeta + e \varphi)
\end{equation}
By multiplying both sides by the electron density $n$ and introducing the pressure $p$ and density of bulk forces $\textbf{f}$
\begin{equation}
\label{pressure}
\nabla p=n\nabla \zeta,~~~~~\textbf{f}=-en\nabla\varphi
\end{equation}
Eq.(\ref{2nd_law}) can be brought into the standard form of the Euler equation of hydrodynamics \cite{LL6} 
\begin{equation}
\label{nr_Euler}
mn\left (\frac{\partial \textbf{u}}{\partial t} +(\textbf{u}\cdot \nabla)\textbf{u}\right )=-\nabla p +\textbf{f}
\end{equation}

\subsubsection{Spectrum of plasma oscillations}

The small density oscillations can be understood by linearizing Eqs.(\ref{continuity}) and (\ref{2nd_law}) about the equilibrium state $n=n_{d}$ and $\textbf{u}=0$.  To first order in $\delta n$ and $\textbf{u}$, the continuity equation (\ref{continuity}) and the equation of motion (\ref{2nd_law}) become \begin{equation}
\label{linearized_continuity}
\frac{\partial \delta n}{\partial t}+ n_{d}\nabla \cdot \textbf{u}=0  
\end{equation}
\begin{equation}
\label{linearized_2nd_law}
m\frac{\partial \textbf{u}}{\partial t}=-\nabla \left (\frac{\partial \zeta}{\partial n_{d}}\delta n+e \varphi\right )
\end{equation}
Differentiating Eq.(\ref{linearized_continuity}) with respect to time and employing Eq.(\ref{linearized_2nd_law}) the velocity field $\textbf{u}$ can be eliminated with the result
\begin{equation}
\label{oscillation_real_space}
\frac{\partial^{2}\delta n}{\partial t^{2}}-\frac{n_{d}}{m}\nabla^{2} \left (\frac{\partial \zeta}{\partial n_{d}}\delta n+e \varphi\right )=0
\end{equation}
Going over to the Fourier representation and employing the Coulomb law $\varphi(\textbf{q})=(e/\kappa)n(\textbf{q})f_{d}(q)$ (Eq.(\ref{potential_Fourier}) with $\varphi_{ext}=0$) turns this into an ordinary differential equation
\begin{equation}
\label{oscillation_Fourier_space}
\frac{d^{2}n(\textbf{q})}{dt^{2}}+\Omega_{d}^{2}(\textbf{q})n(\textbf{q})=0
\end{equation}
This describes a harmonic oscillator with frequency $\Omega_{d}(\textbf{q})$ 
\begin{equation}
\label{nr_spectrum}
\Omega_{d}^{2}(\textbf{q})=s^{2}q^{2}+\frac{n_{d}e^{2}}{\kappa m}q^{2}f_{d}(q)
\end{equation}
\begin{equation}
\label{nr_sound}
s^{2}=\frac{n_{d}}{m}\frac{\partial \zeta}{\partial n_{d}}\equiv \frac{1}{m}\frac{\partial p}{\partial n_{d}}
\end{equation}
where $s$ is the adiabatic speed of sound in the neutral $e=0$ limit \cite{LL6}.  $\Omega_{d}(\textbf{q})$ is the spectrum of the plasma oscillations of the $d$-dimensional electron gas with a parabolic dispersion law.  Eq.(\ref{nr_spectrum}) can be re-written in terms of the static dielectric function $\varepsilon_{d}(\textbf{q})$ (\ref{phi_vs_phi_ext}) as $\Omega_{d}^{2}(\textbf{q})=s^{2}q^{2}\varepsilon_{d}(\textbf{q})$ thus implying that short-wavelength $q \gg q_{s}$ density oscillations are sound-like, $\Omega_{d}(\textbf{q})=sq$, while in the long-wavelength $q \ll q_{s}$ limit the second term in Eq.(\ref{nr_spectrum}) dominates and one has classical plasma waves with $\Omega_{d\neq1}(\textbf{q})\simeq (n_{d} e^{2}/m)^{1/2}q^{(3-d)/2}$ and $\Omega_{1}(\textbf{q})\approx(2n_{1}e^{2}/\kappa m)^{1/2}q \ln^{1/2}(1/qa)$ in agreement with the many-body calculation \cite{DasHwang}.  The well-known $d=3$ version of Eq.(\ref{nr_spectrum}) is  $\Omega_{3}^{2}(\textbf{q})=s^{2}q^{2}+4\pi n_{3}e^{2}/\kappa m$; the discussion of the relationship between hydrodynamic and many-body approaches as well as an analysis of various limiting cases can be found in many places \cite{Pines_Nozieres,Jackson,Fetter,Feynman}.  The $d=2$ version of Eq.(\ref{nr_spectrum}), $\Omega_{2}^{2}(\textbf{q})=s^{2}q^{2}+(2\pi n_{2}e^{2}/\kappa m)q$, was derived by Fetter \cite{Fetter}.  

The macroscopic theory explains how the plasmon spectrum depends on the space dimensionality of the electron system \cite{Dyakonov} which the many-body approach tends to obscure.  In the long-wavelength limit the density gradient term of Eq.(\ref{linearized_2nd_law}) can be neglected, and the rest simplifies to $m\partial \textbf{u}/\partial t=e\textbf{E}$ where $\textbf{E}=-\nabla \varphi$ is the electric field (or the part in plane or along the wire) acting in the electron system.  Combining this  with the time derivative of the linearized continuity equation (\ref{linearized_continuity}) leads to a simplified version of Eq.(\ref{oscillation_real_space})
\begin{equation}
\label{continuity_Gauss}
\frac{\partial^{2}\delta n}{\partial t^{2}}+\frac{n_{d}e}{m}\nabla \cdot \textbf{E}=0
\end{equation}
For $d=3$ one can directly substitute Gauss's law $\nabla \cdot \textbf{E}=4\pi e\delta n/\kappa$ which then predicts that the classical plasma frequency is $\Omega_{3}(\textbf{q}\rightarrow 0)=(4\pi n_{3}e^{2}/\kappa m)^{1/2}$.  This line of reasoning fails for $d=2$ or $d=1$ because $\textbf{E}$ entering Eq.(\ref{continuity_Gauss}) is only a projection of the total three-dimensional electric field onto the $d$-dimensional space of the electron system, while Gauss's law involves all components of the electric field.  In such a situation we proceed by going over to the Fourier representation and directly employing the Coulomb law (\ref{potential_Fourier}) (with $\varphi_{ext}=0$): $(\nabla \cdot \textbf{E})(\textbf{q})=-(\nabla^{2} \varphi)(\textbf{q})=q^{2}\varphi(\textbf{q})=(e/\kappa)q^{2}f_{d}(\textbf{q})n(\textbf{q})$. Then the classical plasma frequency is $\Omega_{d}(\textbf{q}\rightarrow 0)=(n_{d}e^{2}q^{2}f_{d}(q)/\kappa m)^{1/2}\rightarrow (n_{d}e^{2}/\kappa m)^{1/2} q^{(3-d)/2}$ ($d\neq1$) or $\Omega_{1}(\textbf{q}\rightarrow 0)\approx(2n_{1}e^{2}/\kappa m)^{1/2}q \ln^{1/2}(1/qa)$.

\subsubsection{Dynamical dielectric function}
   
Next we evaluate the dynamical (frequency $\omega$ and wave vector $\textbf{q}$ dependent) dielectric function of the system, $\varepsilon_{d}(\omega,\textbf{q})$ \cite{Mahan}.  We substitute $\varphi_{ext}(\textbf{q})=\varphi_{ext}(\omega,\textbf{q})e^{-i\omega t}$ into Eqs.(\ref{potential_Fourier}) and (\ref{oscillation_real_space}), and seek the total potential $\varphi(\textbf{q})$ and density $n(\textbf{q})$ in the form of a driven oscillation: $\varphi(\textbf{q})=\varphi(\omega,\textbf{q})e^{-i\omega t}$ and $n(\textbf{q})=n(\omega,\textbf{q})e^{-i\omega t}$, respectively.  This affects Eq.(\ref{potential_Fourier}) only minimally, adding an $\omega$-dependence to the Fourier transforms of the potentials and the density,
\begin{equation}
\label{potential_Fourier_time}
\varphi(\omega,\textbf{q})=\varphi_{ext}(\omega,\textbf{q})+\frac{e}{\kappa}n(\omega,\textbf{q})f_{d}(q),
\end{equation}
while Eq.(\ref{oscillation_real_space}) becomes            
\begin{equation}
\label{oscillation_time_Fourier}
\frac{\omega^{2}}{q^{2}}n(\omega, \textbf{q})=\frac{n_{d}}{m}\left (\frac{\partial\zeta}{\partial n_{d}}n(\omega,\textbf{q})+e\varphi(\omega,\textbf{q})\right )
\end{equation}
Elimination of $n(\omega,\textbf{q})$ from Eqs.(\ref{potential_Fourier_time}) and (\ref{oscillation_time_Fourier}) establishes that $\varphi(\omega,\textbf{q})=\varphi_{ext}(\omega,\textbf{q})/\varepsilon_{d}(\omega,\textbf{q})$, thus giving the dynamical dielectric function:
\begin{equation}
\label{dyn_diel_function}
\varepsilon_{d}(\omega,\textbf{q})=\frac{\omega^{2}-\Omega_{d}^{2}(\textbf{q})}{\omega^{2}-s^{2}q^{2}}
\end{equation}
The $d=3$ and $d=2$ versions of Eq.(\ref{dyn_diel_function}) were given previously in Refs. \cite{Pines_Nozieres,Barton} and \cite{Fetter}, respectively.  As in the many-body approach, the zero of the dielectric function determines the plasmon spectrum, $\varepsilon_{d}(\Omega_{d},\textbf{q})=0$.  The pole of $\varepsilon_{3}(\omega,\textbf{q})$ at $\omega^{2}= s^{2}q^{2}$ may be thought of as a remnant of particle-hole excitations of the neutral $e=0$ system \cite{Barton};  the same holds for general $d$.  This interpretation is supported by the facts that the pole of the dielectric function is an indicator of the onset of absorption, plasmons are known to decay into particle-hole excitations, and that the hydrodynamic approximation treats all the excitations as density oscillations.  In the $\omega=0$ limit the dielectric function (\ref{dyn_diel_function}) reduces to its static counterpart (\ref{phi_vs_phi_ext}).  In the $\textbf{q}=0$ limit one finds $\varepsilon_{d\neq3}(\omega,0)=1$ and $\varepsilon_{3}(\omega,0)=1-4\pi n_{3}e^{2}/\kappa m\omega^{2}$, a textbook result \cite{Jackson}.  

\subsection{Dispersion law with limiting velocity}

While the continuity equation (\ref{continuity}) still applies to the Dirac plasma,  Euler's equation (\ref{nr_Euler}) does not apply because it is based on Newton's second law, which does not hold for massless excitations (\ref{dispersion_law}).  The formalism appropriate for treatment of this case is developed in relativistic hydrodynamics \cite{LL6} which we now adopt (substituting the speed of light $c$ with the Fermi velocity $v_{F}$).  We emphasize that the electron liquids in materials with limiting velocity are not Lorentz-invariant with $v_{F}$ playing a role of the speed of light $c$.        

The central object of the theory is the energy-momentum tensor of the liquid
\begin{equation}
\label{en_mom_tensor}
T^{ik}=wu^{i}u^{k}-pg^{ik}, ~~~u^{i}=\gamma \left (1, \frac{\textbf{u}}{v_{F}}\right )
\end{equation}
where $w$ is the heat function density, $u^{i}$ is the velocity vector, $\gamma=(1-u^{2}/v_{F}^{2})^{-1/2}$, and $g^{ik}$ is the metric tensor with components:  $g^{00}=1$, $g^{ ii}=-1$ ($i\neq0$), and $g^{ik}=0$ otherwise.  The equations of motion of the liquid are the statements that the divergence of the energy-momentum tensor $T^{ik}$ is due to the bulk force density $f^{i}$ \cite{LL2,LL6}:
\begin{equation}
\label{eqs_of_motion}
\frac{\partial T^{ik}}{\partial x^{k}}=f^{i}, ~~~f^{i}=\left (\frac{\textbf{f}\cdot\textbf{u}}{v_{F}},\textbf{\textbf{f}}\right )
\end{equation} 
where $x^{i}=(v_{F}t, \textbf{r})$ is the position vector.  Evaluation of the temporal $i=0$ component of Eq.(\ref{eqs_of_motion}) gives the energy balance equation
\begin{equation}
\label{en_balance}
\frac{\partial}{\partial t}(w\gamma^{2}-p)+\nabla \cdot (w\gamma^{2}\textbf{u})=\textbf{f}\cdot\textbf{u}
\end{equation}
which can be employed to bring the spatial $i\neq0$ component of Eq.(\ref{eqs_of_motion}) into a useful form
\begin{equation}
\label{rel_Euler}
\frac{w\gamma^{2}}{v_{F}^{2}}\left (\frac{\partial \textbf{u}}{\partial t} +(\textbf{u}\cdot\nabla)\textbf{u}\right )=-\nabla p+ \textbf{f}-\frac{\textbf{u}}{v_{F}^{2}}\frac{\partial p}{\partial t}-\frac{\textbf{u}(\textbf{f}\cdot\textbf{u})}{v_{F}^{2}}
\end{equation}   
Eqs.(\ref{en_balance}) and (\ref{rel_Euler}) were given previously \cite{Fogler}. The latter is a generalization of the standard Euler equation to the case of a generic dispersion law, which also makes it possible to precisely state the range of applicability of Eq.(\ref{nr_Euler}).  For slow motions, $u\ll v_{F}$, one can set $\gamma=1$ and the last two terms in the right-hand side of (\ref{rel_Euler}) can be neglected.  Since the heat function density $w$ is the sum of the energy density $\epsilon$ and pressure $p$, $w=\epsilon+p$ \cite{LL6}, the chemical potential is the derivative of the energy density, $\zeta(n)=\partial \epsilon/\partial n$, and the pressure and chemical potential are related as $\partial p/\partial n=n\partial \zeta/\partial n$, Eq.(\ref{pressure}), one has $w=n\zeta(n)$, and Eq.(\ref{rel_Euler}) becomes 
\begin{equation}
\label{rel_Euler_slow_limit}
\frac{\zeta(n)}{v_{F}^{2}}n\left (\frac{\partial \textbf{u}}{\partial t} +(\textbf{u}\cdot\nabla)\textbf{u}\right )=-\nabla p+ \textbf{f}, ~~~u\ll v_{F}
\end{equation}   
This resembles the Euler equation (\ref{nr_Euler}) except that the counterpart of mass $m$ is now the density-dependent combination $\zeta(n)/v_{F}^{2}$ that parallels Einstein's equivalence relationship $E=mc^{2}$ between mass and energy.  The limit of a parabolic spectrum, Eq.(\ref{nr_Euler}), can now be recovered by writing $\zeta=mv_{F}^{2}+\zeta_{int}(n)$ and neglecting in the left-hand side of (\ref{rel_Euler_slow_limit}) the "internal" part of the chemical potential $\zeta_{int}$ (which does contribute into the pressure gradient, Eq.(\ref{pressure}), in the right-hand side).

The analysis of small oscillations for a generic dispersion law differs from the parabolic case only in that the role of the Euler equation (\ref{nr_Euler}) is played by its generalization (\ref{rel_Euler}).  Since the oscillations are perturbations about the $n=n_{d}$, $\textbf{u}=0$ state, the outcome can be written out without additional calculations by replacing the electron mass $m$ by $\zeta(n_{d})/v_{F}^{2}$ whenever the mass is encountered in the previously given formulas \cite{cyclotron}.  This has no effect on the static screening properties accumulated in Eqs.(\ref{Fourier_of_Coulomb})-(\ref{static_diel_function}), but modifies the dynamical predictions.  Specifically, the spectrum of the plasma oscillations (\ref{nr_spectrum}) generalizes to      
\begin{equation}
\label{rel_spectrum}
\Omega_{d}^{2}(\textbf{q})=s^{2}q^{2}+\frac{n_{d}e^{2}v_{F}^{2}}{\kappa \zeta(n_{d})}q^{2}f_{d}(q)
\end{equation}
with the expression for the speed of sound (\ref{nr_sound}) modified to
\begin{equation}
\label{re_sound}
s^{2}=v_{F}^{2}\left (\frac{\partial p}{\partial \epsilon}\right )_{n=n_{d}}
\end{equation}
This is a counterpart of the expression for the speed of sound in a relativistic liquid with $v_{F}$ playing the role of the speed of light $c$ \cite{LL6}.  For the Dirac plasma the equation of state is $p=\epsilon/d$ and the speed of sound is $s=v_{F}/\sqrt{d}$.  

Eqs.(\ref{rel_spectrum}) and (\ref{re_sound}) in combination with the expression for the dynamical dielectric function (\ref{dyn_diel_function}) accumulate the basic information regarding screening and plasma oscillations in a generic $d$-dimensional electron gas in the hydrodynamic approximation.  Even though a modification of the standard hydrodynamics \cite{LL6} is necessary to accommodate for the possibility of the Dirac dispersion law (\ref{dispersion_law}), at no point was quantum mechanics employed:  to order $q^{2}$ quantum-mechanical effects do not explicitly enter the expression for $\Omega_{d}^{2}(\textbf{q})$ (\ref{rel_spectrum}).  In the long-wavelength limit the second term in (\ref{rel_spectrum}) dominates and one finds 
\begin{equation}
\label{rel_spectrum_long_waves}
\Omega_{d\neq1}(\textbf{q}\rightarrow 0)\simeq  \left (\frac{n_{d}e^{2}v_{F}^{2}}{\kappa \zeta(n_{d})}\right )^{1/2}q^{(3-d)/2}
\end{equation}   
\begin{equation}
\label{rel_1d_spectrum_long_waves}
\Omega_{1}(\textbf{q}\rightarrow 0)\simeq  \left (\frac{n_{1}e^{2}v_{F}^{2}}{\kappa \zeta(n_{1})}\right )^{1/2}q\ln^{1/2}\left (\frac{1}{qa}\right )
\end{equation}
One can make Planck's constant $\hbar$ "reappear" in these equations by looking at a model for the electron gas.  For example, at zero temperature the chemical potential of the electron gas (\ref{dispersion_law}) of degeneracy $g$ and density $n_{d}$, is $\zeta(n_{d})\simeq \hbar v_{F}(n_{d}/g)^{1/d}$ (neglecting the effects of exchange and correlation), and then the dependence of the plasmon frequency on the density and Planck's constant $\hbar$ is given by 
\begin{equation}
\label{density_dependence}
\Omega_{d}(\textbf{q}\rightarrow 0)\propto \left (\frac{n_{d}^{1-1/d}e^{2}v_{F}g^{1/d}}{\kappa \hbar}\right )^{1/2} 
\end{equation} 
which agrees with and explains the results of the many-body calculation for this model \cite{DasHwang}.  We hasten to mention that in graphene's case ($d=2$) the equivalence of hydrodynamic and many-body plasmon spectra of this model has already been noticed \cite{Svintsov}.  We emphasize that the conclusions (\ref{rel_spectrum}) and (\ref{re_sound}) are not limited to zero temperature or a specific model of the electron gas:  they are completely general with the system-specific information encoded in the equation of state $\zeta(n_{d})$.  The hydrodynamic approach is inherently limited to long-wavelength low frequency phenomena.  It complements the powerful but somewhat abstract methods of the many-body theory, highlighting the macroscopic origin of the effects in question. 

\section{Acknowledgements}
  
We thank D. Svintsov for informing us of Ref. \cite{Svintsov}.

\end{document}